\documentclass[preprint,secnumarabic,amssymb, nobibnotes, aps, prb,superscriptaddress]{revtex4-1}
\usepackage{graphicx}

\begin{document}

\title{Tunable Thermal Transport and Thermal Rectification in Strained Graphene Nanoribbons}
\author{K.G.S.H. Gunawardana}
\email[Electronic address:]{harsha@ou.edu}
\affiliation{Homer L. Dodge Department of Physics and Astronomy, Center for Semiconductor Physics in Nanostructures, The University of Oklahoma, Norman, Oklahoma 73069, USA}
\author{Kieran Mullen}
\affiliation{Homer L. Dodge Department of Physics and Astronomy, Center for Semiconductor Physics in Nanostructures, The University of Oklahoma, Norman, Oklahoma 73069, USA}
\author{Jiuning Hu}
\affiliation{Birk Nanotechnology Center and  School of Electrical and Computer Engineering, Purdue University, West Lafayette, Indiana 47907, USA}
\author{Yong P. Chen}
\affiliation{Birk Nanotechnology Center and Department of Physics, Purdue University, West Lafayette, Indiana 47907, USA}
\affiliation{Birk Nanotechnology Center and School of Electrical and Computer Engineering, Purdue University, West Lafayette, Indiana 47907, USA}
\author{Xiulin Ruan}
\affiliation{ Birk Nanotechnology Center and School of Mechanical Engineering, Purdue University, West Lafayette, Indiana 47907, USA}
\begin{abstract}

Using  molecular dynamics(MD) simulations, we study thermal transport in graphene nanoribbons (GNR) subjected to uniform uniaxial and nonuniform strain fields. We predict significant thermal rectification (over
$70\%$) in a rectangular armchair GNR by applying a transverse force
asymmetrically. The heat flux is larger from  the less stressed
region to the  more stressed region. 
Furthermore, we develop a theoretical framework based
on the non-equilibrium thermodynamics to discuss when
thermal rectification under a stress gradient can occur.
We conclude with a discussion of details relevant to experiment.

\end{abstract}

\maketitle
% =============================================================================
\section{Introduction}
Operation of nanoscale thermal devices, mainly relies on the
tunability of phonon transport by external means. 
Thermal devices, such as thermal rectifiers\cite{model}, thermal transistors\cite{trans} and thermal memories\cite{memo} are a new class
of devices, whose operation is driven by the temperature gradients.
These devices will have useful applications not only in thermal circuits\cite{Phononics} but also in nanoscale thermal management and thermo-electric applications.

A thermal rectifier, in which the thermal current is larger in one
direction than in the opposite, is one of the most fundamental
thermal devices to be realized experimentally in nanoscale.
 Thermal rectification has been demonstrated using molecular dynamics
 simulations (MD)
in asymmetric systems and is discussed as resulting
 from an interplay between structural asymmetry and lattice
 anharmonicity \cite{model,Asy1,Asy2}.
Recently, this phenomenon has been observed experimentally in
asymmetrically mass-loaded carbon and boron nitride nanotubes
\cite{chang}.  Moreover, thermal rectification has been predicted
in triangular and trapezoidal shaped graphene
nanoribbons(GNRs)\cite{Asy1,Asy2}.  In this letter, we introduce a
method based on strain engineering as an experimentally feasible
approach to realize significant ($>70\%$) 
thermal rectification in GNRs. The potentially
real-time tunability of the thermal rectification is a distinctive
advantage of this approach.

Graphene is a promising material for nanoscale applications  due
to its exceptional electronic \cite{elecp}, thermal \cite{therp}
and mechanical \cite{mechp} properties.  It has been shown recently
that the electronic properties of graphene can be tuned significantly
by engineering strain.  For instance, strain could produce a
pseudo-magnetic field \cite{guinea,tony,levy} which affects the
electronic properties of graphene.  Furthermore, strain could induce
semiconducting properties on metallic GNRs by opening a gap in the
electronic band structure \cite{semi}.  However, strain induced
tuning of thermal transport properties of graphene has been less
studied. The effect of  uniform uniaxial strains on the thermal
conductivity ($\kappa$) of graphene has been studied recently by
MD simulations and the reported effect is substantial in tuning
$\kappa$ \cite{straintc}.

When increasing the uniaxial tensile strain (arrows in fig.\ref{fig:AGNRforce}a), we observe broadening the gap in the phonon dispersion of armchair GNR (AGNR) whereas the effect on the phonon dispersion of zigzag GNR (ZGNR) is mild.
The calculated $\kappa$ of AGNR also reduce considerably
when increasing the tensile strain compared to that of ZGNR.
We predict thermal rectification on an asymmetrically stressed
rectangular GNR.
 In our system, asymmetric stress is achieved by exerting a transverse
 force ($F_y$) on the top and bottom edge atoms of part (for example,
 the right half) of the AGNR as illustrated in fig.\ref{fig:AGNRforce}b.
We observe that the thermal current is larger from less stressed
region to the more stressed region (left to right) than that in the
opposite direction (right to left).  Furthermore, we develop a
theoretical frame work based on the non-equilibrium thermodynamics,
to model thermal rectification in the
presence of a stress gradient.

\section{Simulation procedure}
\begin{figure}\centering \includegraphics[width=3.5in]{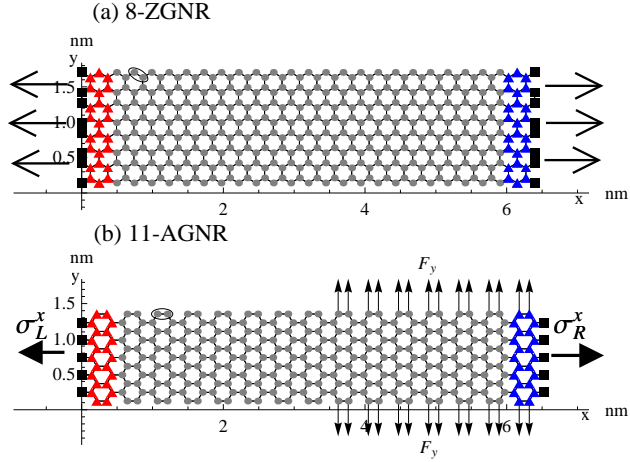}
 \caption{{\protect\small Schematic of the $8$-ZGNR (a) and  $11$-AGNR (b). The triangles represent the thermostated atoms and the full squares are the fixed atoms. The horizontal arrows in (a) shows the direction of uniaxial tensile strain. In (b) the vertical arrows show the applied
 constant force $F_y$ on
 the top and bottom atoms of the right half of the
 AGNR. The horizontal arrows in (b) represent the stress developed in length ($x$) direction ($\sigma_R^x$ and $\sigma_L^x$) near the heat baths. 
  }} \label{fig:AGNRforce} \end{figure}
 Figure \ref{fig:AGNRforce} shows a schematic of  $n$-ZGNR (Zigzag GNR) and $n$-AGNR structures simulated. The carbon-carbon bond length is $1.42  \rm \AA$ in the absence of strains.
The  $n$ refer to the number of carbon dimers ( a dimer is shown by the oval) in width \cite{gillen}.
  The first and the last columns of atoms are fixed and the adjacent  three columns of atoms of the $n$-ZGNR and four columns of atoms of the $n$-AGNR are coupled to the Nos$\acute{e}$ -Hoover thermostats.
In the MD simulation,  a second generation Brenner potential
 \cite{brenner} is employed to describe the carbon-carbon interactions.
 The equations of motion are integrated using the third-order predictor-corrector method.
The time step  is $0.5$ fs and the total simulation time is $5$ ns ($10^7$ time steps). 
 The  temperatures of the left and right heat baths (HBs) are set to
 $T_L$ and $T_R$ respectively. The temperature difference 
 $\left ( T_L -T_R \right ) = 2 \alpha T$, where $\alpha$ determines the temperature bias and the average temperature  $T=(T_L +T_R)/2$.
 The net heat flux ($J$) was calculated by the power delivered by the heat baths \cite{powerhb}. The thermal conductivity ($\kappa$) of the system is  calculated according to Fourier's law, $ J = \kappa  w  h  \nabla T$,
where $w$ and $h$ are width and van der Waals
thickness ($h=0.335$ nm)  of the GNR. 

\section{Thermal transport under uniform strain fields}
\begin{figure}[bt]
 \centering
\includegraphics[width=3.00in]{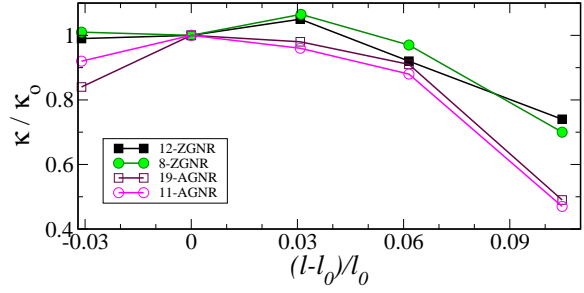} 
  \caption{{\protect\small Normalized thermal conductivity ($\kappa/\kappa_0 $) as a function of uniaxial strain $(l-l_0)/l_0$ along the transport direction(x) calculated at $300$ K.
 }}
 \label{fig:TCvS} \end{figure} 
 Figure \ref{fig:TCvS} shows the variation of  $\kappa/\kappa_0$ of ZGNRs and AGNRs of two different widths as a function of strain  ranging from $-0.03$ to $0.12$. The $\kappa_0$ is the thermal conductivity of unstrained nanoribbon.
 The horizontal arrows in fig.\ref{fig:AGNRforce}a show the direction of uniaxial tensile strain on both nanoribbons.
 The strain is given by $(l-l_0)/l_0$, where $l$ is the stretched or compressed length and $l_0$ is the initial length of the sample.
 Strain is applied by slowly moving the fixed atoms on one side (right side) at a rate of $\pm 1.0\times 10^{-6} \rm \AA$ per time step at $300$ K.
 The calculated values of the thermal conductivity ($\kappa_0$) of unstrained $8$-ZGNR, $12$-ZGNR, $11$-AGNR and $19$-AGNR are $5360$ W/mK, $5500$ W/mK,$3300$ W/mK and $3600$ W/mK respectively.
The variation of $\kappa$ for ZGNRs in the range of strain $0-0.1$ is about $30\%$ from the unstrained values.
 The reduction of $\kappa$ at a strain of $0.1$
 is about $55\%$  for $11$-AGNR and $19$-AGNR.
 \begin{figure}[bt] \centering
 \includegraphics[width=3.25in]{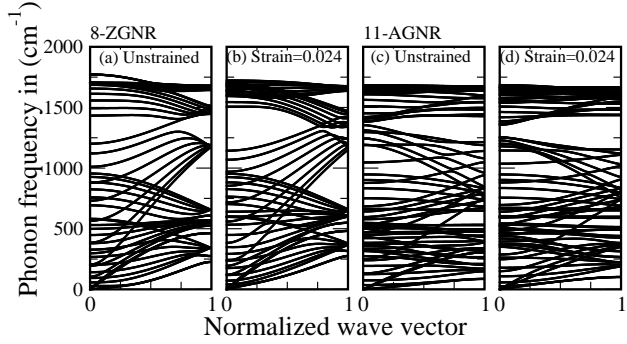} 
   \caption{{\protect\small Phonon dispersion of 8-ZGNR and 11-AGNR under uniaxial tensile strains. The horizontal axis is the normalized wave vector projected on to the transport direction(x).}}
 \label{phononGNR} \end{figure} 
Figure \ref{phononGNR} shows the calculated phonon dispersions of $8$-ZGNR and $11$-AGNR along the transport direction(x) of unstrained  nanoribbons and at a strain of $0.024$. 
When increasing the uniaxial tensile strain, we do not observe a significant  alteration of dispersion curves of the $8$-ZGNR (fig.\ref{phononGNR} a,b). 
However, we observe  an increased of the phonon velocity ($d\omega /dk$) of some modes in the low ($0$-$600$ cm$^{-1}$) and high ($1250$-$1600$ cm${-1}$) frequencies which could be the reason for the increment of the $\kappa$ in small strains.
In $11$-AGNR we observe 
broadening the narrow gap at $1350$ cm$^{-1}$ in the phonon dispersion. At a strain of $0.024$, we observe a gap in the frequency range: $1250-1350$ cm$^{-1}$ (fig.\ref{phononGNR}d). The gap further broaden and moves toward low frequencies when increasing the uniaxial strain.  
This could be the reasons for the relatively larger reduction of  $\kappa$ of AGNR with increasing strain.   

\section{Thermal transport under nonuniform  strain fields}  
 In many of the theoretical studies on thermal transport, the stress along the nanoribbon is considered to be uniform and uniaxial.
We consider a situation that the stress
 at the left and right edges of the nanoribbon are  $\sigma_L$ and
 $\sigma_R$ such that $\sigma_L \neq \sigma_R$. 
 In the intermediate region it is assumed to vary smoothly.
 This kind of stress profile could possibly be realized experimentally using the differential thermal expansion of graphene grown on structured substrates \cite{levy,Bao,raman1} or through externally applied strain \cite{other}.
\begin{figure}[bt]\centering
\includegraphics[width=3.0in]{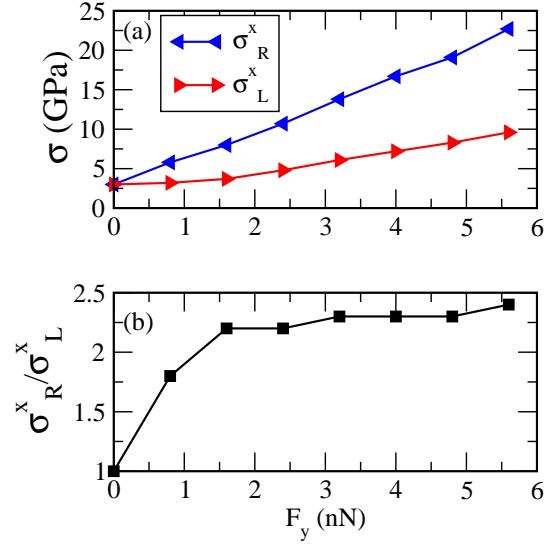}
\caption{{\protect\small (a) Variation of  stresses ($\sigma^x_R$ and $\sigma^x_L$) on the $11$-AGNR as a function of $F_y$. These are the stresses in the center part of the nanoribbon near the right and left heat baths.  (b) The ratio $\sigma_R^x/\sigma_L^x$  as a function of $F_y$. The units of $F_y$ is nano Newton (nN).}}
\label{forces}
\end{figure} 
 We achieve such an asymmetric stress profile in our simulation by transversely stressing a half
of the GNR. For example, we apply a constant force $F_y$ on the atoms of the top and bottom edges of
the right half of the $11$- AGNR  as  depicted in fig.\ref{fig:AGNRforce}b.
The tensile stress developed across the $y$ direction of the right half is denoted by $\sigma^y_R$.
  This lateral stress produces a tensile
 strain along the y direction and,
 hence a compressive strain along the $x$-direction  due to the Poisson
 contraction.  Since the atoms in the left and right edges are fixed, 
 this compressive strain also results in tensile stress in the
x-direction.
The stress in the $x$-direction near the left and right HB is denoted by $\sigma_L^x$ and $\sigma_R^x$.
This is calculated from the time average of the forces on the fixed atoms at the left and right edges(excluding the corner atoms).
Figure \ref{forces}a shows the variation of $\sigma_L^x$ and
$\sigma_R^x$ on $F_y$ calculated at $T=300$ K.
  The $\sigma_L^x$ and $\sigma_R^x$  are tensile, and
 $\sigma_L^x$  is smaller than the $\sigma_R^x$ while  both are  found
 to be increasing with $F_y$.

\subsection*{Thermal rectification}
 \begin{figure}[bt] \centering
 \includegraphics[width=3.00in]{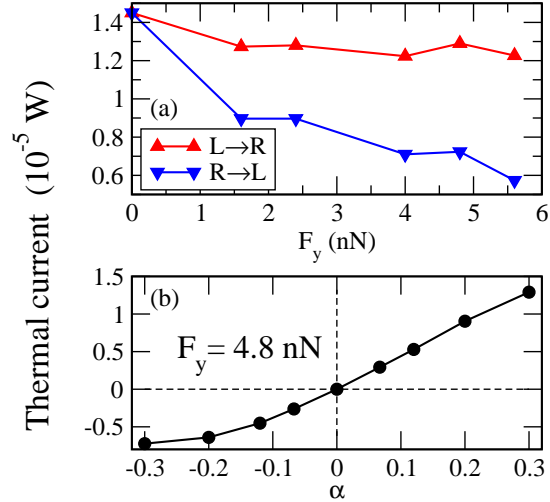} 
  \caption{{ \protect\small (a)  Thermal current from left to right and right to left  as a function of $F_y$ at $T=300$ K and $|\alpha|=0.3$, and (b) Variation of thermal current with the temperature bias ($\alpha$) at $F_y=4.8$ nN.}}
 \label{TRec} \end{figure} 
In fig.\ref{TRec}a, the thermal current from left to right ($+\alpha$)
and right to left ($-\alpha$) are plotted against the applied force
$F_y$, when $|\alpha| =0.3$ and $T=300$ K. The thermal current from
left to right ($J_{L \rightarrow R}$) shows very small decrement,
where as the $J_{R \rightarrow L}$ decreases significantly when
$F_y$ is increased. Thus, the $J_{L \rightarrow R}$ is considerably
larger than $J_{R \rightarrow L}$ whenever $F_y > 0$,
 marking the existence of thermal rectification in this system.
 
The tensile stress and strain can affect the thermal transport 
by distorting the lattice to both alter the characteristic vibrational
frequencies and the degree of anharmonicity.\cite{tensor}
 Anharmonicity is essential: \emph{no amount of geometric or parameter
asymmetry can produce thermal rectification.} Without anharmonicity,
the system can be analyzed via normal modes, and time reversal symmetry
will require that transmission amplitudes  from left to right are the
same as right to left.
 We attribute the observed thermal rectification
 to the strain induced asymmetry of the vibrational frequencies and
 the lattice anharmonicity.  Such an asymmetry of the vibrational
 properties leads to a local variation of $\kappa$.  It can be
 shown that the thermal
 conductivity must be a function of both position and temperature
 to rectify the thermal
 current.\cite{supple}  

We can invoke the framework of non-equilibrium thermodynamics
to describe this effect.
 The change of the entropy per unit volume of a solid due to
 the applied stress ($\tau_\alpha$)
  can be expressed\cite{Wallce}, 
$ds=\frac{du}{T}-\frac{1}{T} \sum_\alpha \tau_\alpha d\eta_\alpha$,
where $du$ is the
 change of internal energy density, 
$\eta_\alpha$ is the strain and $\alpha\in{[xx,yy,xy]}$.
 From this we can deduce the rate of production of entropy\cite{callen}, 
\begin{equation}
\dot{s}= \partial_i \left(\frac{1}{T}\right) J_i^{(u)} + 
\frac{1}{T} %\sum_\alpha
  \partial_i \left(\tau_\alpha \right) J^{(\eta_\alpha)}_i \label{EnProd}
 \end{equation}
where $J^{(u)}_i$ and $J^{(\eta_\alpha)}$ are the $i$-th components
of the energy and the strain
currents, and repeated indices are summed over.
In general the currents 
are a function of the intensive
parameters (T and $\tau$) as well as the  affinities\cite{callen}
($\nabla\frac{1}{T}$ and $\nabla\tau$).  Thus the heat
current, $\vec J^{(Q)}$, can be expanded in its most general form:

\small
\begin{eqnarray}\nonumber
 J_i^{(Q)}=
L^{(Q)}_{i,j}\, \partial_j \frac{1}{T}
 +  \frac{L^{(\eta_\alpha)}_{i,j}}{T} \, \partial_j\tau_\alpha
 +L^{(QQ)}_{i,j,k}\, \partial_j \frac{1}{T} \partial_k \frac{1}{T} 
 + \\
 \frac{L^{(Q\eta_\alpha)}_{i,j,k}}{T}
\partial_j \frac{1}{T}\partial_k \tau_\alpha +
\frac{L^{(\eta_\alpha\eta_\beta)}_{i,j,k}}{T^2}\partial_j \tau_\alpha 
\partial_k \tau_\beta \label{PS2}
\end{eqnarray} \normalsize 
In solids, barring plastic deformation,  there is no heat current
in steady state solely due to $\nabla\tau_i$.
Thus, the coefficients 
$L^{(\eta_\alpha) }=L^{(\eta_\alpha\eta_\beta)}=0$.  
Moreover, in symmetric
systems there is no thermal rectification, which implies $L^{(QQ)}=0$.
If we further assume only gradients in the $x$-direction 
and stress in the $y$-direction
 the above equation can be reduced to, 
\small \begin{equation}
J_x^{(Q)}=L_{x,x}^{(Q)}\partial_x \frac{1}{T} +
\frac{L^{(Q\eta_{yy})}_{x,x,x} }{T}\partial_x \frac{1}{T}\partial_x \tau_{yy},
\label{hflux} \end{equation} \normalsize
 which leads to different thermal currents when switching the
 sign of the stress gradient ($\nabla\tau$).
 For the case of $F_y=4.8$ nN, the calculated kinetic coefficients are
 $L^{(Q)}_{x,x}=1.7\times10^8$ WK/m and $L^{(Q\eta_{yy})}_{x,x,x}=0.34$ WK$^2$/GPa.

 \begin{figure}[bt] \centering
 \includegraphics[width=3.0in]{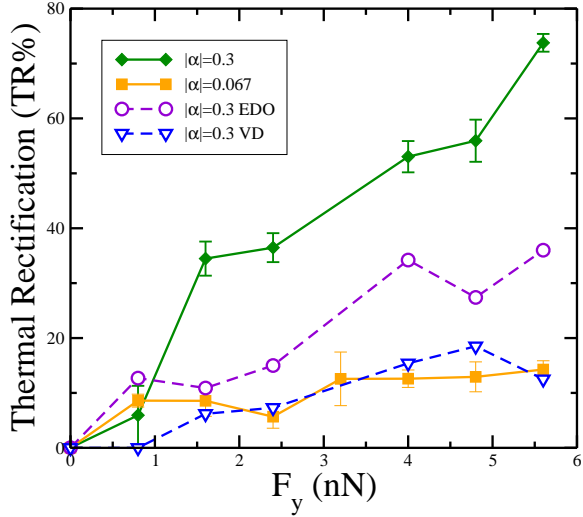} 
 \caption{{ \protect\small  Thermal rectification factor(TR) at $T=300$ K as a function of $F_y$. The open circles are in the presence of edge disorder(EDO) and the open triangles are in the presence of vacancy defects(VD). }}
\label{TRfac} \end{figure} 
The distinctive property of this approach is the tunability of the
thermal rectification by an asymmetrically applied force, $F_y$.
We show in fig.\ref{TRfac} the  thermal rectification factor (TR)
 as a function of $F_y$ calculated at $T=300$ K.
The TR  is defined as,
\begin{equation}
 TR = 2 \frac{(J_{L \rightarrow R} - J_{R
\rightarrow L})}{(J_{L \rightarrow R} +
 J_{R \rightarrow L})} \times 100\%.
 \end{equation}
We observe an increment of the TR when increasing the $F_y$.  The
increment of TR with $F_y$ is more prominent at higher biases
($\alpha=0.3$).  As shown in the fig.\ref{TRec}b, the $J_{L \rightarrow
R}$  increases almost linearly with the bias ($\alpha$), whereas
$J_{R \rightarrow L}$ increases nonlinearly giving rise to a larger
TR.  This behavior suggests the contribution of higher order terms
(beyond the second order) to eq.\ref{PS2}.  The nonlinear transport
is an essential property in realizing thermal rectification as in
an electronic diode.  In our system the nonlinear transport is
prominent at larger $F_y$'s ($>0.8$ nN), where we observe a clear
bias dependence of TR.  The maximum TR we observed  is about $73\%$
which occurred at $\alpha=0.3$ and $F_y=5.6$ nN.

 The right side of the AGNR is subjected to a biaxial stress whereas
 the left side only has an uniaxial stress.
It is evident that the asymmetry of the on axis stresses ($\sigma_R^x$
and $\sigma_L^x$) is important in determining the TR.  We
achieve a significant TR whenever the ratio $\sigma_R^x/\sigma_L^x
\gtrsim 2.0$ (see fig.\ref{forces}b).
 When we move the laterally stressed window towards the center, the
 TR reduced considerably in our simulation.  This is a reasonable
 observation, because the asymmetry of the system is reduced.
For instance, when we move the laterally stressed window eight
columns towards the center (keeping the number of atoms that $F_y$
is applied constant) the TR decrease to $5\%$ at $\alpha=0.3$ and
$F_y=4.8$ nN. In this situation, the ratio of  stresses reduces to
$1.5$ which is close to the ratio at $F_y=0.8$ nN where the observed
TR is about $6\%$. 
When the length of the nanoribbon is increased, the strain gradient reduces and results a reduction of thermal rectification (implies from the eq.\ref{hflux}).
However, by increasing the length of the AGNR and also
the number of atoms that $F_y$ is applied, we could increase the
strain gradient and observe a larger TR.

Graphene commonly possesses edge disorder which significantly
degrade the thermal properties.  In fig.\ref{TRfac} the open
circles shows the variation of TR in the presence of edge disorder.
Since the simulated AGNRs are very small, we introduce only about a
$4\%$( percentage of number of edge atoms removed)  edge disorder.
We observe moderate reduction of TR, down to a value  
$\sim 30\%$.  The effect of edge disorder can be reduced by increasing
the width of the nanoribbon.  However, the presence of vacancy defects
significantly reduces the TR. The open triangles in fig.\ref{TRfac} is for
$0.6\%$ vacancy defects distributed evenly in the nanoribbon.

Experimentally, the lateral force on the right half of the AGNR can
be applied by coupling to a substrate. Our simulation
fixed the atoms to which the forces were applied. This situation
also produces larger thermal rectification(over $100\%$) at higher
biases and the direction of the maximum thermal current is same as
before.  This effect can be understood by the argument
in ref.\cite{supple}.  Consequently the observable net thermal
rectification could be even higher due to the asymmetric coupling
between the GNR and the substrate. 

Electronic transport of heat will occur in parallel to the phonon 
conduction, but is not so large that it dwarfs the phonon channel
considered here.  In addition, AGNRs
shows semiconducting characteristics and their energy gap can be tuned
with the strain\cite{semi}. Thus, the electronic contribution to the thermal
transport should not be crucial. 
Electron-phonon interactions could lead to
processes that undermine the thermal rectification.  We believe
that their contribution to the thermal rectification is also minimal
since the nanoribbon is semiconducting. In addition, the long
electronic coherence length in graphene indicates that electron-phonon
interactions should not be significant.
In experiments, these effects can
be further minimized by electronically gating the sample.

Finally, we discuss the stability of the C-C bonds in  graphene under the lateral forces.
We do not observe any rupture of bonds within the applied range of $F_y$ ($0 - 5.6$ nN).
In our simulation the estimated maximum force in the C-C bond direction  is about $6.4$ nN, which is correspond to a lateral stress of $\sigma^y_R=90$ Gpa.
In a recent experiment, it has been found that the intrinsic strength of a single layer graphene is about $130$ GPa assuming the van der Waals thickness of graphene \cite{mechp}.
When the $F_y$ is increased to $7.2$ nN, we observe rupturing of some bonds near the fixed atoms in the right side of the AGNR.
At this point, the maximum force in the C-C bond direction is about $8.0$ nN. 
Thus, the forces required to realize thermal rectification by our method, are realistic and in an experimentally feasible range.

\section{Conclusion}
In conclusion, we study the thermal transport properties of strained GNRs using MD simulations. 
We demonstrate that  the thermal rectification can be realized by engineering the stress on a rectangular AGNR. 
We have found that the heat transport is favorable  from the less stressed region to the more stressed region. We also found that edge defects and
vacancies only partially suppress this rectification. 
The major advantage of this approach is that the thermal rectification can be tuned from no rectification state  to over $70\%$ in real-time by applying a mechanical force.

\begin{acknowledgements}
This project was supported in part by the US National Science Foundation
under Grant~\mbox{MRSEC DMR-0080054}. JH and YPC acknowledge the support by NRI-MIND.
\end{acknowledgements}

%\clearpage
% -----------------------------------------------------------------------------

\end{document}